\documentclass[conference]{IEEEtran}
\IEEEoverridecommandlockouts
\usepackage{url}

\usepackage{cite}
\usepackage{amsmath,amssymb,amsfonts}
\usepackage{algorithmic}
\usepackage{graphicx}
\usepackage{textcomp}
\usepackage{xcolor}

\usepackage{lipsum}  % 示例文本
\usepackage{fancyhdr}

\usepackage{float}
\usepackage{diagbox}
\usepackage{comment}
\def\BibTeX{{\rm B\kern-.05em{\sc i\kern-.025em b}\kern-.08em
    T\kern-.1667em\lower.7ex\hbox{E}\kern-.125emX}}
\begin{document}

\title{DEthna: Accurate Ethereum Network Topology  Discovery with Marked Transactions%\\
%\thanks{The corresponding author is Shengli Zhang. The research is partially supported by National Natural Science Foundation of China (62171291), Shenzhen Key Research Project (JSGG20220831095603007, JCYJ20220818100810023, JCYJ20220818101609021) and Science and Technology Program (JCYJ20210324094609027), Australian Research Council (ARC) Discovery Project (DP210101723), and Hong Kong RGC Research Impact Fund (No. R5060-19, No. R5034-18), Areas of Excellence Scheme (AoE/E-601/22-R) and  General Research Fund (No. 152203/20E, 152244/21E, 152169/22E, 152228/23E).}
\thanks{The corresponding author is Shengli Zhang (zsl@szu.edu.cn). The research is supported in part by the National Natural Science Foundation of China (62171291), in part by the Shenzhen Key Research Project (JSGG20220831095603007, JCYJ20220818100810023, JCYJ20220818101609021), in part by the Shenzhen Science and Technology Program (JCYJ20210324094609027), in part by the Australian Research Council (ARC) Discovery Project (DP210101723), and in part by the Hong Kong RGC Research Impact Fund (No. R5060-19, No. R5034-18), Areas of Excellence Scheme (AoE/E-601/22-R) and  General Research Fund (No. 152203/20E, 152244/21E, 152169/22E, 152228/23E).}
}

% \and
% \and

% \author{\IEEEauthorblockN{Chonghe Zhao\IEEEauthorrefmark{2}, Yipeng Zhou\IEEEauthorrefmark{3}, Shengli Zhang\IEEEauthorrefmark{2}, Taotao Wang\IEEEauthorrefmark{2}, Quan Z. Sheng\IEEEauthorrefmark{3}, Song Guo\IEEEauthorrefmark{4}}
% \IEEEauthorblockA{\IEEEauthorrefmark{2}College of
% Electronic and Information Engineering, Shenzhen University, Shenzhen, China\\
% \IEEEauthorrefmark{3}School of Computing, Macquarie University, Sydney, Australia\\
% \IEEEauthorrefmark{4}Department of Computer Science and Engineering, The Hong Kong University of Science and Technology, Hong Kong\\
% Email: zhaochonghe\_szu@163.com, yipeng.zhou@mq.edu.au, \{zsl, ttwang\}@szu.edu.cn\\ michael.sheng@mq.edu.au, songguo@cse.ust.hk}
% }
\DeclareRobustCommand*{\IEEEauthorrefmark}[1]{%
    \raisebox{0pt}[0pt][0pt]{\textsuperscript{\footnotesize\ensuremath{#1}}}}
\author{\IEEEauthorblockN{Chonghe Zhao\IEEEauthorrefmark{1,2}, Yipeng Zhou\IEEEauthorrefmark{2}, Shengli Zhang\IEEEauthorrefmark{1}, Taotao Wang\IEEEauthorrefmark{1}, Quan Z. Sheng\IEEEauthorrefmark{2}, Song Guo\IEEEauthorrefmark{3}}
\IEEEauthorblockA{\IEEEauthorrefmark{1}College of
Electronic and Information Engineering, Shenzhen University, Shenzhen, China\\
\IEEEauthorrefmark{2}School of Computing, Macquarie University, Sydney, Australia\\
\IEEEauthorrefmark{3}Department of Computer Science and Engineering, The Hong Kong University of Science and Technology, Hong Kong\\
Email: zhaochonghe\_szu@163.com, yipeng.zhou@mq.edu.au, \{zsl, ttwang\}@szu.edu.cn\\ michael.sheng@mq.edu.au, songguo@cse.ust.hk}
}

%\IEEEauthorrefmark{1}Corresponding author: Shengli Zhang

\maketitle

\begin{abstract}
In Ethereum, the ledger exchanges messages along an underlying Peer-to-Peer (P2P) network to reach consistency. 
Understanding the underlying network topology of Ethereum is crucial for network optimization, security and scalability.  However, the accurate discovery of Ethereum network topology is non-trivial due to its deliberately designed security mechanism. Consequently,  
existing measuring schemes %the Ethereum network topology %based on UDP-based node discovery mechanism and fake transactions 
cannot accurately infer the Ethereum network topology with a low cost. To address this challenge, we propose the \underline{D}istributed \underline{Eth}ereum \underline{N}etwork \underline{A}nalyzer (DEthna) tool, which can accurately and efficiently measure the Ethereum network topology. 
In DEthna,  a novel  parallel measurement model is proposed that can generate marked transactions to  infer link connections based on  the transaction replacement and propagation mechanism in Ethereum. %marked transactions broadcast  inherent
%To control the cost to generate transactions for measurement, a parallel measurement scheme is devised. 
Moreover, a workload offloading scheme is designed so that DEthna can be deployed on multiple distributed probing nodes so as to  measure a large-scale Ethereum network at a low cost. % with nodes distributed in different ISPs (Internet Service Providers). 
We run DEthna on Goerli (the most popular Ethereum test network) to evaluate its capability in discovering network topology. The experimental results demonstrate that DEthna significantly outperforms the state-of-the-art baselines.
Based on DEthna,  we further analyze characteristics of the  Ethereum network revealing that there exist more than 50\%  low-degree Ethereum nodes that weaken the network robustness. %are behind NAT (network address translation) or firewalls  to unveil that the ``long tail” broadcast latencies, and hence  weakening network robustness. 
%Ethereum relies on the underlying Peer-to-Peer network to exchange messages to maintain the consistency of the ledger. Understanding its underlying network topology is a crucial factor for network optimization, security, and scalability. However, it is difficult to discover the network topology due to the deliberately secure design from onlookers. Meanwhile, the current measurement schemes for Ethereum network topology are limited by their inability to accurately infer the links among nodes and the fact that they miss detecting a large number of nodes by relying on a single probe node. This paper presents the distributed Ethereum Network Analyzer (DEthna), a tool that measures the network topology of the Ethereum blockchain and analyze its network characteristics. In DEthna, we first propose a novel link inference model that exploits the mechanism of transaction replacement and propagation to infer the connections accurately. Then, we design a distributed measurement architecture in which multiple modified Ethereum-probing nodes adopt the link inference model to discover the network topology in parallel. We run DEthna on Goerli (the most popular Ethereum test network) to verify its availability, discover the network topology, and analyze the network characteristics. The experimental results demonstrate that DEthna outperforms the current measurement schemes in terms of both precision/recall and network size. With discovered Ethereum topology, we find that more than half of nodes behind NAT or firewall incur the “long tail” latencies and weaken the network robustness. 
\end{abstract}


\section{Introduction}\label{intro}
Blockchain, which was first proposed by Nakamoto in Bitcoin \cite{b1},  is a prominent technology  integrating the advances of cryptography, distributed computing and Peer-to-Peer (P2P) communication.  Following Bitcoin, various blockchain systems  emerged to tackle different technical challenges. Ethereum \cite{b2} is one of the most successful and influential blockchains, which can support the logic computation of various applications (\emph{e.g.}, Decentralized Finance \cite{b3}, Non-Fungible Token \cite{b4} and Metaverse \cite{b5}) via smart contracts. 

As a fully distributed Internet system, Ethereum  exploits a P2P network to build its communication infrastructure. A new node participates in the system by discovering and connecting with other nodes in the P2P network. All nodes send and receive pending transactions and confirmed blocks over the P2P network to maintain the consistency of the data recorded on the blockchain. 
As the wide adoption of Ethereum, understanding its  network topology becomes extremely important for network optimization, security, and scalability. For example, Zhao \emph{et al.} \cite{b43} applied the knowledge of network topology measured in \cite{b16} to verify its ``Bodyless Block Propagation" scheme for speeding up block propagation in blockchain networks adopting Proof-of-Work (PoW) consensus algorithm \cite{shi2022pooling}.
%{\bf YP: from these two examples, I just saw the vulnerability of Ethereum by leaking the network topology. It seems these are not explaining the importance of netowork topology for network optimization, security and scalability. Please find new examples to support this statement. For example. this information can improve network robustness? we need some positive examples. } 

However, Ethereum is vulnerable to malicious attacks if its network topology information is unveiled. 
For instance, an attacker aware of the network topology can link blocks to originating nodes, and the decentralization of Ethereum can be compromised by performing Denial-of-Service (DoS) attacks or eclipsing attacks on these nodes \cite{b19,b7,b8,b9}. 
In \cite{b10},  a broadcast ``advantage” can be set up if the network topology knowledge is available, which enables an attacker to pull off double-spending attacks. To prevent potential attacks using network topology information, Ethereum is designed to conceal its network topology from onlookers.   

Despite the invisibility of the underlying network topology in Ethereum, significant research efforts have been dedicated to indirectly measuring the topological characteristics of the Ethereum network, such as \cite{b11,b12,b13,b14}.
%Although there have been several works  that aim to, 
Unfortunately, none of them can accurately discover the network topology of a large-scale Ethereum with a low cost. In prior works \cite{b11,b12,b13},  discovery table based methods were designed  to infer connections between nodes. For a specific node, remote nodes stored in its discovery table are inferred as its neighbors. Yet,  the accuracy of such methods is inferior   because  each node only selects a small subset of remote nodes in its discovery table to establish connections.
In \cite{b14}, Li \emph{et al.} proposed to measure the Ethereum topology by  issuing a few marked transactions plus more than 5,000 invalid transactions to infer the connection between two nodes. Although this approach can achieve a high measurement accuracy, it floods the network with an excessive number of invalid transactions, which is 
 prohibited by countermeasure used to prevent network congestion \cite{b15}. In addition,  its single probing node architecture makes the measurement of large networks inefficient. % It cannot efficiently discover the topology of a large-scale network because it only deploys a  single probing node for measurementhas two significant deficiencies: 1) It is constrained caused by a large number of invalid transactions. \textcolor{blue}{2) It cannot efficiently discover the topology of a large-scale network because it only deploys a  single probing node for measurement.}
%due to its heavy communication cost. 
\begin{table*}[h]

\caption{Methods and Weaknesses of the works aiming to measure network topology of public blockchain systems}
\begin{tabular}{|l|l|l|l|}

\hline

\textbf{Reference}     & \textbf{Methodology}                                                               & \textbf{Blockchain} & \textbf{Weakness}                                                                                                               \\ \hline
\cite{b18}           & Exploit random node discovery mechanism                              & Bitcoin    & Restricted by Countermeasure \cite{b22, b23}                                                                           \\ \hline
\cite{b19}           & Perform timing analysis of transaction propagation                   & Bitcoin    & Low precision and recall                                                                                               \\ \hline
\cite{b20, b21}      & Issue and monitor marked transactions                               & Bitcoin    & \begin{tabular}[l]{@{}l@{}}\cite{b20}: Low recall; \\ \cite{b21}: Restricted by Countermeasure \cite{b45}\end{tabular} \\ \hline
\cite{b11, b12, b13} & Exploit the node discovery mechanism based on Kademlia DHT protocol      & Ethereum   & Low precision and recall                                                                                               \\ \hline
\cite{b16}           & Analyze blocks and transactions propagation                           & Ethereum   & Fail to exactly infer each link between  nodes                                                                                 \\ \hline
\cite{b14}           & Issue marked and fake transactions, and monitor marked transactions & Ethereum   & \begin{tabular}[l]{@{}l@{}}Harmful for network performance and  not \\  scalable in a large-scale network\end{tabular}                              \\ \hline
\end{tabular}
\label{compareWeakness}

\end{table*}

To overcome deficiencies in the existing methods, we propose  a \underline{D}istributed \underline{Eth}ereum \underline{N}etwork \underline{A}nalyzer (DEthna) to accurately and efficiently discover the network topology of a large-scale Ethereum network. 
%by deploying multiple modified Ethereum nodes to 
Compared with prior works, DEthna has two advantages: 1) The number of transactions required per link inference is low;  2) DEthna is scalable by offloading the measuring workload among multiple probing nodes. 
%DEthna  contains a link inference model, in which a probing node exploits real transactions to accurately discover connection links between Ethereum nodes without incurring invalid transactions. 
%2) 
More specifically, DEthna exploits the transaction replacement and propagation mechanism in Ethereum to infer multi-links by only generating a certain number of real transactions without generating invalid transactions. Based on our design, a proposal \cite{bxx} is submitted to the Ethereum community by fixing a defect in the standard Ethereum software, which has been accepted and committed to the software.
To improve DEthna's scalability, a workload offloading scheme is designed to deploy multiple distributed probing nodes 
%for collective measurement across a diverse Ethereum network spanning various ISPs (Internet Service Providers).
to jointly measure a large-scale Ethereum network in which Ethereum nodes may reside in various ISPs (Internet Service Providers). 
%a distributed measurement architecture to infer connections among nodes precisely to discover the network topology and analyze the network characteristics. 

We implement DEthna and the state-of-the-art baselines by modifying the standard Ethereum software Geth \cite{b28} and deploy them on Goerli, the most popular Ethereum test network. %As a comparison of measurement effect, we also implement the existing measurement methods and deploy them on Goerli, and the 
We conducted extensive experiments. The experimental results demonstrate that DEthna significantly outperforms baselines in terms of  link inference accuracy. Based on our measurement study, we further analyze  the characteristics of the Ethereum network  to unveil that  more than 50\%  are low-degree nodes that are probably behind NAT (network address translation) or firewall  incurring  the ``long tail” communication latency, and hence weakening the network robustness. % against the targeted attack. It implies that the robustness of the Ethereum network can be substantially improved by  reducing the number of  NAT nodes  or connecting  them directly. 

The rest of this paper is organized as follows. Section \ref{rel} discusses  related works. Section \ref{background} presents the background of the  transactions and node discovery mechanism in Ethereum. The link inference model and our distributed measurement architecture are elaborated in Section \ref{link} and Section \ref{linkParallel}, respectively.   
The experimental results  together with network characteristics analysis are presented in Section \ref{experiment} before we finally conclude our work in Section \ref{conclusion}.  

\section{Related Work} \label{rel}

 {Topology knowledge  plays a vital role in optimizing, securing, and scaling the blockchain network. In particular, Bitcoin and Ethereum are the two most representative public blockchain networks. Accurately measuring their topology has attracted tremendous research attention in recent works.}

%Network topology is the basis for the security and performance of public blockchain systems. The knowledge of network topology crucial to understanding its robustness against various attacks, but also 
%Measuring network topology is indispensable for analyzing network characteristics of published blockchains, including network size, decentralization, robustness, and so on. In particular,  for Bitcoin and Ethereum, the two most representative blockchain systems, measuring their network topology is a challenging but essential problem that has been  attempted by numerous prior works.
%{\bf YP: In the first paragraph, it is not clear why measuring topology is important, and specifically what we can do with the measured topology. If this topology information can be only used for designing attack algorithms, I believe this measuring study is not very important. }
%{\bf YP: please use cite command, it is better if all references are put in a separate bib file. } 

Bitcoin employs a random node discovery mechanism to establish an unstructured network \cite{simblock}.  A gossip message broadcast protocol is used to exchange blocks and transactions between nodes \cite{b7}. %Based on its node discovery mechanism and  message broadcast protocol, 
Several prior works \cite{b18,b19,b20,b21} have contributed to measuring Bitcoin network topology and analyzing its network characteristics. Miller \emph{et al.} \cite{b18} proposed a Bitcoin network analyzer (AddressProbe) to infer each connection between nodes for reconstructing the network topology by using timestamps recorded in node address messages. But countermeasures \cite{b22,b23} updated in Bitcoin Core nodes make AddressProbe not feasible anymore. Neudecker \emph{et al.} \cite{b19} performed a timing analysis of transaction propagation to measure Bitcoin network topology with inferior precision and recall (both around 40\%). To improve  \cite{b19}, Grundmann \emph{et al.} \cite{b20} presented two inference methods by exploiting the accumulation of multiple transactions and the behavior of dropping double-spending transactions. Delgado-Segura \emph{et al.} \cite{b21} introduced a novel technique (TxProbe) for reconstructing network topology by ``orphaned” transactions \cite{test}.

Measuring the network topology of Ethereum is also challenging but essential. There are two differences between Ethereum and Bitcoin making the aforementioned measurement methods invalid in Ethereum. Firstly, Ethereum adopts the K-bucket data structure in the Kademlia DHT protocol \cite{b11} to discover network nodes and maintain node address information. Secondly, the account model of transactions used in Ethereum is very different from the UTXO transaction model in Bitcoin \cite{b25}. In view of these differences, \cite{b11,b12,b13,b14,b16} devised new methods to measure the network topology of Ethereum. The studies in \cite{b11, b12, b13} reconstructed the network topology by  node information stored in the K-bucket data structure, though their measurement accuracy is unsatisfactory. %they cannot precisely infer each link among nodes. 
The work \cite{b16} exploited the Ethereum network degree distribution and the number of hops for transmitting messages by using the block and transaction propagation protocol. However, it cannot exactly infer each link between nodes, and thus the inferred entire network topology is inaccurate. 
The work \cite{b14} inferred each link between two nodes to discover the network topology by issuing marked transactions. Yet, this method is harmful to the network system performance with a large number of actual links that are missed for detection. 
 % and provide the experimental results in Section \ref{experiment}. 

To better understand our contribution, we  summarize typical blockchain network measurement studies with their methodologies and weaknesses  in Table \ref{compareWeakness}. %Existing methods developed for Bitcoin are not applicable to measure Ethereum  %their poor performance and the limitation of specific countermeasures. Moreover, these methods are not compatible with Ethereum 
%due to the fundamental differences between Bitcoin and Ethereum. Therefore, dedicated  works have been contributed to  measuring the network topology of Ethereum. Yet, they still cannot achieve high accuracy in discovering the network topology with a low cost. % without negatively impacting the network. 
From the table, we can find that no existing work can accurately and efficiently discover Ethereum topology, which is to be addressed by our work. %  To verify that our method outperforms these baselines proposed in \cite{b11, b12, b13, b14}, we have re-implemented these methods for performance comparison   with ours.}
%{\bf YP: after reading the related work section, it is not clear what challenges we have solved, what weaknesses in existing works have been overcome by us. Is it possible to summarize the challenges of measuring topology and weaknesses of existing works, that have been solved by us. In this way, it will be clear to show our contribution. }
% \usepackage[normalem]{ulem}
% \usepackage[normalem]{ulem}

\section{Ethereum Background}\label{background}
%In this section, we introduce the background of Ethereum before we discuss the design of DEthna. 
Before we introduce the design of DEthna, we provide the background of Ethereum in this section.
\subsection{Transaction Fields}\label{txsF}
%In essence, a blockchain is  a distributed ledger. Blockchain nodes modify data stored in the ledger in a decentralized but mutually consistent manner by running transactions in the block. To present, there are two blockchain transaction models \cite{b25}, the Unspent Transaction Outputs (UTxO) model of Bitcoin and the account model of Ethereum. The UTxO model is stateless supporting the token ownership transfer. In contrast, the account model adopted by Ethereum is stateful supporting the execution of generic programs, a.k.a, smart contracts,  on the blockchain. %Ethereum adopts the account model with smart contract functionality. %to extend blockchain applications. 

In Ethereum, a transaction, to be executed by Ethereum Virtual Machine (EVM), binds a sender account to a receiver account \cite{graph}. Each computational step in EVM is priced in the unit of \emph{gas}.  Field values  in each transaction specify the implementation of the cryptocurrency ether (ETH) transfer and execution of smart contracts. In particular,  four key fields in Ethereum are used  for designing DEthna  after the effectiveness of the  Improvement Proposal EIP-1559 \cite{b26} in 2021:%The price the sender wishes to pay per unit of gas is GasPrice. The transaction fee is equal to GasPrice multiplied by the total gas required to execute the transaction. }
%Before EIP-1559 is adopted in Ethereum, all transaction fee will be rewarded to miners. But for EIP-1559 \cite{b26} is adopted in current Ethereum, part of the transaction fee will be burned, and the remaining transaction fee will be rewarded to miners.} 

%Based on the account model, an Ethereum transaction binds a sender account to a receiver account. It is executed by consuming gas through the Ethereum Virtual Machine (EVM), with each computational step in the EVM priced in units of gas. contains several fields to implement the cryptocurrency ether (ETH) transfer and execute smart contracts, and four key fields are used to design our DEthna:

\begin{itemize}
    \item $nonce$: It is a monotonically increasing counter indicating the number of transactions  issued from a sender account. Transactions from the same sender account must be packaged and executed in an ascending order of $nonce$.
    \item $data$: It stores the code related to  smart contracts or any content meaningful to the sender. For example, the transaction issued by DEthna includes the word ``DEthna" in $data$ to label the transaction. 
    \item $gasTipCap$ $(g_t)$: It is the maximum price of  consumed gas that a sender wishes to incentivize proposers/miners to include the transaction in the next block.
    \item $gasFeeCap$ $(g_f)$: It is the maximum price per gas unit that a sender would like to pay for a particular transaction. 
    %that the sender is willing to pay for the transaction.
    %incentivizes proposer/miner to prioritize the transaction.
    %\item \textcolor{blue}{$GAS\;Price$: Each transaction is executed to consume GAS by Ethereum Virtual Machine (EVM), and every computational step in EVM is priced in units of GAS. The price the sender wishes to pay per unit of GAS is $GAS$ $Price$. The transaction fee is equal to $GAS$ $Price$ multiplied by the total GAS required to execute the transaction. That is, the larger $GAS\;Price$, the larger transaction fees. Additionally, after EIP-1559 \cite{b26} is adopted in Ethereum, $GAS\;Price$ for each transaction dynamically changes according to the size of the last block.}
\end{itemize}

According to  EIP-1559, the effective price per gas that miners can get is  $\min \{g_f-b_f,g_t\}$ by including a transaction in a block. Here, $b_f$ is  $baseFee$  indicating the minimum price per gas unit burned by a transaction, which is recorded in the block header and adjusted dynamically according to network conditions. In other words, the transaction fee has two parts:  Part 1 is paid to miners which is equal to the effective price per gas ($\min \{g_f-b_f,g_t\}$) multiplied by the total gas consumed to execute the transaction; Part 2 removed from circulation is equal to $b_f$ multiplied by the total gas consumed. Usually, a marked transaction in DEthna \cite{b29} consumes 24,152 gas.

\subsection{Transaction Processing}
Transactions are propagated between  Ethereum nodes before they are confirmed and recorded in a block. Each  transaction before confirmed is processed by an Ethereum node by the  following rules:

%When an Ethereum node receives one or multiple unconfirmed transactions from other Ethereum nodes over the network, it will process transactions according to the following rules: 

\textbf{Transaction Store:} Each Ethereum node maintains a transaction pool consisting of a Pending submodule and a Queue submodule to store received unconfirmed transactions. According to the order of $nonce$, transactions are classified as pending and future transactions.  Pending transactions are with continuous $nonces$ (\emph{i.e.}, transactions issued from the same account contain a consecutive sequence of $nonces$) and stored in the Pending submodule. Future transactions are with discontinuous $nonces$ and stored in the Queue submodule. For example,  if an account has issued $n$ transactions, $TX_1$ issued by this account is regarded as a pending transaction when the $nonce$ of $TX_1$  is $n$ and other transactions with $nonces$ equal to $1, 2,\dots n-1$ have been received.
Otherwise, if the $nonce$ of a transaction $TX_2$ issued by the same account is $n+1$, and the Ethereum node has not received $TX_1$ with $nonce=n$ yet, $TX_2$ is regarded as a  future transaction.  $TX_2$ will not be transformed into a pending transaction until $TX_1$ is received. 

% To avoid the spam attack, the space of the transaction pool is limited and is set at node startup. Usually, for the node adopting the most popular Ethereum software Geth, the default size of Pending is 5,120, and the default size of Queue is 1,024. When the number of transactions reaches to 5,120+1,024 (full), new transactions with high effective price will evict the old ones with lower effective price, and readjust the position of transactions in Pending and Queue according to their $nonces$. 
 
\textbf{Transaction Propagation Mechanism:} When an Ethereum node receives a pending transaction, \emph{e.g.,} $TX_1$, it stores $TX_1$ in the Pending submodule and then forwards $TX_1$ to its neighbor nodes. When an Ethereum node receives a future transaction, \emph{e.g.,} $TX_2$, $TX_2$ will be stored in the Queue submodule without further forwarding. Only when all transactions with a smaller $nonce$ arrive, $TX_2$ can be transformed into a pending transaction for further forwarding. 

\textbf{Transaction Replacement Mechanism:} If the effective price of a transaction is too low,  it cannot be packaged into a new block quickly. In this case, the sender may issue a new transaction with the same $nonce$ but a higher effective price to replace the old one. The relation of effective price between these two transactions must satisfy: 
\begin{equation}
    ({b_{t2}} - {b_{t1}})/{b_{t1}} \ge \alpha,
    	\label{predictionpre}
\end{equation}
where $b_{t1}$ and $b_{t2}$ are the old  and new effective price, respectively. The replacement rate $\alpha$ is a constant fixed at node startup. Otherwise, if Eq.~\eqref{predictionpre} cannot be satisfied, the new transaction  will be discarded.%{\bf YP: what happen if eq 1 is not satisfied?  does it mean replacement will not happen? If no replacement, how to do with the new transaction with bt2?}

\section{Link Inference Model in Ethereum}\label{link}

In this section, we introduce the link inference model in DEthna, and  analyze its failure probability and overhead cost. % present a single link inference model that utilizes marked transactions to infer a link between two Ethereum nodes, building upon the Ethereum background discussed in the previous section. We then extend this model to propose a multi-link inference model, which aims to reduce transaction fees associated with issuing marked transactions. Finally, we conduct an analysis of failed inference cases.

\subsection{Single Link Inference Model}\label{linkaaa}

To ease the understanding of the link inference model, we firstly discuss the simplest scenario to infer a single  link between two particular nodes.
Suppose that there is  a measurement node $M$ that can be controlled by DEthna, and our objective is to infer whether the  target node $A$ is directly linked with another target node $B$ for exchanging  blockchain messages. Let $C$ denote all the  rest nodes in the Ethereum network. %The measurement node $M$ focuses on inferring a single link between target node $A$ and target node $B$, that is, whether $A$ and $B$ are connected in the Ethereum network to .
Note that the measurement node $M$ is  a full Ethereum node that can monitor unconfirmed transactions from the network and store them in its local transaction pool. 

\textbf{Marked Transactions:} Marked transactions are real transactions that are generated by node $M$ dedicated  for discovering links. To infer the link between nodes $A$ and $B$, node $M$ needs to generate four marked transactions (denoted by $TX_M$, $TX_A$, $TX_B$ and $TX_C$) with a well defined relationship. These transactions should be sent to Ethereum nodes in a deliberately designed order, which is further explained as follows: %The relationships between four marked transactions are explained in the following:
\begin{figure*}[tp]
\centering
\includegraphics[width=\linewidth,height=6.0cm]{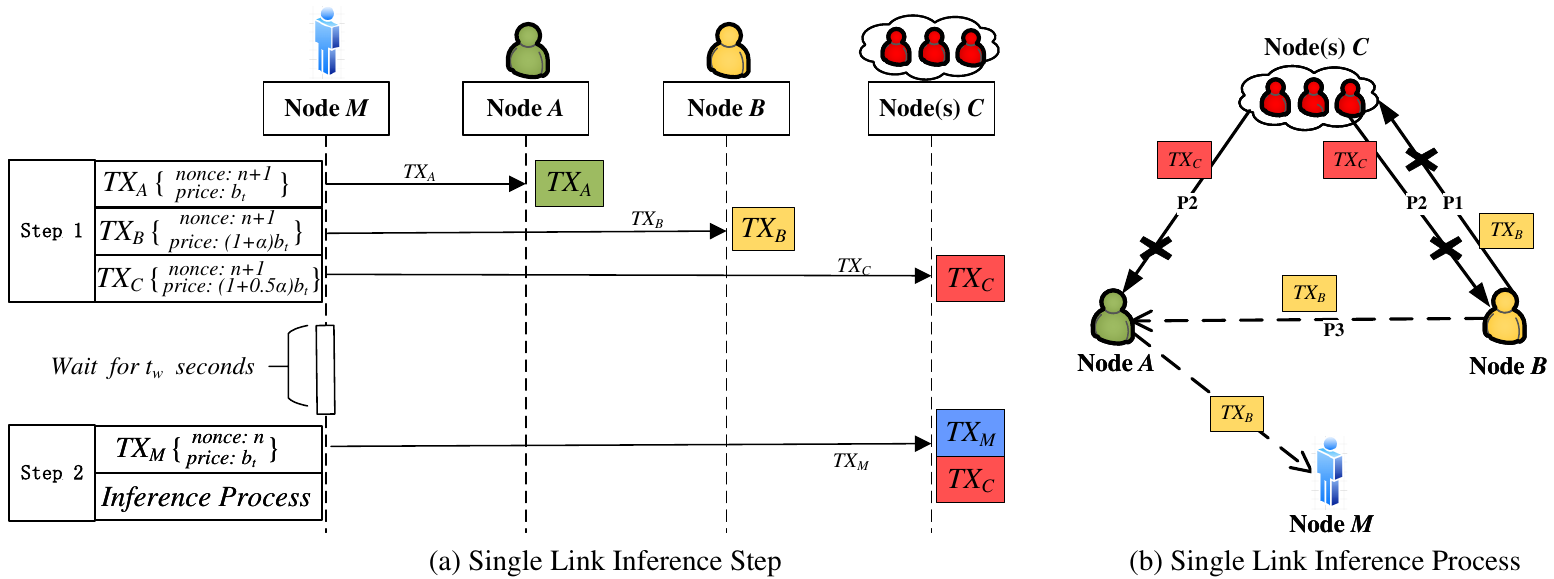}
\caption{Single Link inference model to infer a link between nodes $A$ and $B$: (a) $TX_A$, $TX_B$ and $TX_C$ with nonce $n+1$ are regarded as future transactions and not forwarded over the network before $TX_M$ with nonce $n$ is sent out; (b) $TX_A$, $TX_B$ obeying \textbf{P1, P2\&P3}, and $TX_C$ are transformed to pending transactions and forwarded after $TX_M$ arrives, and $TX_B$ can arrive at node $A$ only when there is a link between nodes $A$ and $B$.}
\label{fig1}
\end{figure*}
\begin{itemize}
\item 	All four marked transactions are issued by the same account of node $M$. Without loss of generality, we suppose that the account has issued $n-1$ historical transactions; 
\item  	Each marked transaction like \cite{b29} records the IDs of nodes $A$ and $B$ in the data field;
\item 	 $TX_M$ is set  with the effective price $b_t$ and the nonce $n$;
\item   $TX_A$ is set with the effective price $b_t$ and the nonce $n+1$; 
\item  	 $TX_B$ is set with the effective price $(1+ \alpha )b_t$ and the nonce $n+1$;
\item    $TX_C$ is set with the effective price $(1+0.5 \alpha)b_t$ and the nonce $n+1$.
\end{itemize}

\noindent\textbf{Inference Steps:} The core idea of our link inference model is to isolate the exchange of transaction $TX_B$  between node $B$ and other nodes except $A$, and $TX_B$ can only be further forwarded by $A$ to $M$. 
It is conducted in two steps. In the first step, node $M$ distributes $TX_A$, $TX_B$ and $TX_C$  to nodes $A$, $B$ and $C$, respectively,  as future transactions, which will not be exchanged between nodes. In the second step, node $M$ sends out $TX_M$ which can trigger the exchange  of  $TX_A$, $TX_B$ and $TX_C$ between nodes. Then, node $M$ infers the link between $A$ and $B$ by  monitoring  whether  $TX_B$ can be forwarded back to node $M$. Each step in more detail is presented as follows.%A more detailed description of each step is presented as follows.  

    \textbf{Step 1:} Node $M$ sends $TX_A$, $TX_B$ and $TX_C$ to nodes $A$, $B$ and $C$, respectively. Since  they do not  have $TX_M$ with nonce $n$, these transactions with nonce $n+1$ are regarded as future transactions that will be only stored in  their Queues submodules without being further forwarded. % by  $A$ to any other nodes.
  %\textbf{Step 2:} 
 % Similarly, node $M$ sends $TX_B$ and $TX_C$ to  nodes $B$ and $C$, respectively.  Here, the nonce of both $TX_B$ and $TX_C$ is $n+1$, and thus they are regarded as future transactions which will not be exchanged between nodes. % with nonce $n+1$ is regarded as a future transaction to be stored in  $B$’s Queue as well.  Similarly,  node $M$ sends $TX_C$ to all other node(s) $C$. $TX_C$ with nonce $n+1$ is also regarded as a future transaction to be  stored in $C$’s Queue. %And node(s) $C$ do not continue to forward $TX_C$ to other nodes. 
  
 Since there may exist  thousands of nodes $C$, $M$ waits for $t_{w}$ seconds after the first step to ensure all these marked transactions can successfully reach nodes $A$, $B$ and $C$. %receives $TX_B$, and all node(s) $C$ receive $TX_C$, since usually there are hundreds of thousands of nodes over the Ethereum network and the nodes cannot forward these future transactions.

\textbf{Step 2:} Node $M$ sends $TX_M$ to nodes $C$ to trigger the exchange of $TX_A$, $TX_B$ and $TX_C$ in the system. 
In other words, once nodes $C$ receive $TX_M$, $TX_C$ becomes a pending transaction, which will be forwarded to nodes $A$ and $B$ along with the forwarding of $TX_M$. %to make them forward over the network. As shown in Fig.\ref{fig1}(b), when node(s) $C$ receives $TX_M$, $TX_C$ will be transformed to a pending transaction and moved to Pending, and then forwarded along with $TX_M$ to node $A$ and node $B$. For $TX_A$, $TX_B$, and $TX_C$ with the same nonce $n+1$, 
%\alpha$ of $TX_A$’s effective price $b_t$ and $TX_B$’s effective price $(1+\alpha)b_t$.
Then, node $M$ monitors the system and infers that the direct link  between  $A$ and $B$ exists if  $TX_B$  is forwarded back to node $M$. 

The effectiveness of the link inference model is guaranteed by the following properties for forwarding $TX_B$:
\begin{itemize}
    \item \textbf{Property 1 (P1).} $TX_B$ cannot reach nodes $C$  given that $TX_M$  is sent by $M$ to  nodes $C$ first before  $TX_B$ becomes pending  and the effective price of $TX_B$ is not enough to replace $TX_C$. 
    \item \textbf{Property 2 (P2).} $TX_C$ cannot replace $TX_A$ or $TX_B$ because  its effective price $(1+0.5 \alpha)b_t$ is lower than the threshold (\emph{i.e.}, $(1+\alpha)b_t$) for replacement.

    \item \textbf{Property 3 (P3).} $TX_B$ can exactly replace $TX_A$ given its price is above the threshold to replace $TX_A$, and thus $A$ can further forward $TX_B$ to other nodes. 
    
\end{itemize}
Based on \textbf{P1, P2\&P3}, $TX_B$ will not be forwarded to $M$ unless a direct link between $A$ and $B$ exists so that $TX_B$ can be forwarded to $M$ along the path $B\rightarrow A\rightarrow M$. Meanwhile, our inference model can guarantee that  $TX_B$ cannot be forwarded to node $A$ via node(s) $C$ according to \textbf{P1}.

To better understand its effectiveness, we show the link inference process in Fig.~\ref{fig1}(a). In particular, the inference step is presented in Fig.~\ref{fig1}(b), in which \textbf{P1\&P2} guarantee the forward isolation of $TX_A$, $TX_B$ and $TX_C$, while \textbf{P3} guarantees that $TX_B$ can be forwarded to $M$ if a direct link connecting $A$ and $B$ exists.

%node $M$ infers the link between $A$ and $B$ with five steps  based on four marked transactions.
  %   \textbf{Step 5:} Node $M$ infers a link if it receives $TX_B$ from node $A$. Because both $TX_A$ and $TX_B$ are isolated by node(s) $C$ and $TX_A$ is isolated by node $B$, and only $TX_B$ can be stored in node $A$’s Pending by replacing $TX_A$ ($TX_B$’s effective price $(1+\alpha)b_t$ is equal to $\alpha$ of $TX_A$’s effective price $b_t$).  However,  $TX_C$ will not be further forwarded because threshold to replace $TX_A$ and $TX_B$. by nodes $A$ and $B$ cannot replace $TX_A$ and $TX_B$, and it is thereby isolated by node $A$ and node $B$ (i.e., only $TX_M$ arrives at node $A$ and node $B$), since $TX_C$’s  As $TX_M$ is received by node $A$ and node $B$, both $TX_A$ and $TX_B$ are transformed to the pending transactions and moved to Pending, and then forwarded along with $TX_M$ to other nodes.(not sending $TX_M$ to node A and node $B$ can improve the precision of our model, and the detailed discussion can be saw in Section \ref{link}-C).

  %(\emph{i.e.}, a 5-step inference task):

{\bf Inference Cost Analysis.} In our link inference model, it takes 4 marked transactions to infer the existence of a single link. In comparison, it takes more than 5,120 transactions for measuring a link in \cite{b14}. Thereby,  the communication overhead of DEthna is much lighter. However, in DEthna, $TX_M$ and one of $TX_A$, $TX_B$ and $TX_C$ will be recorded into  the blockchain by measuring a single link, which costs transaction fees. As discussed in Section \ref{txsF}, transaction fees are determined by the variable  $baseFee$ ($b_f$). During off-peak hours when there are fewer transactions in the system, $b_f$ tends to be smaller. Thus, to minimize the price cost of DEthna, it is suggested to execute DEthna during off-peak hours. %Note that only one of $TX_A$, $TX_B$ and $TX_C$ will consume price because they have the same nonce, and one of them is recorded into the blockchain. 

\subsection{Multi-link Inference Model}

Since $TX_M$ consumes transaction fees for measuring a single link, we further propose the multi-link inference model, which can measure multi-links with a single $TX_M$. Thereby, we can discover the Ethereum network topology with a lower fee cost by implementing the multi-link inference model.  

%In the single link inference model, each 5-step inference task requires four marked transactions and results in two marked transactions ($TX_M$ with nonce $n$ and one of $TX_A$, $TX_B$, and $TX_C$ with nonce $n+1$) being packaged in the block. In other words, a single $TX_M$ is used to  infer a link only. If there are $l$ nodes in the network, node $M$ needs to perform $\frac{{l(l - 1)}}{2}$ link inferences to infer all possible links, thereby it needs to pay transaction fees for $l(l - 1)$ marked transactions to be packaged in the block. To reduce transaction fees, we extend the single link inference model and propose a multi-link inference model. 

Suppose that our objective is to consume a single $TX_M$ to measure $K$ links. %be used for parallel inference of $k$ links, enabling cost savings in terms of transaction fees.
The design principle is very similar to that of the single link inference model. We generate $TX_A^k$, $TX_B^k$ and $TX_C^k$ with effective prices 
 $b_t$, ${\rm{(1 + }}\alpha ){b_t}$, and ${\rm{(1 + 0}}{\rm{.5}}\alpha ){b_t}$, respectively,   for measuring link $k$ connecting node $A_k$ and node $B_k$.  Their nonce value is set to $n+k$. 
The multi-link inference model also has two steps, which are briefly described as follows by emphasizing its difference from the single link inference model.

\textbf{Step 1.}
Node $M$ distributes $TX_A^k$, $TX_B^k$ and $TX_C^k$  to nodes $A_k$, $B_k$ and $C_k$, respectively, for $k=1,\dots, K$. Note that all these transactions with nonce $n+K$ are future transactions, which will not be exchanged between nodes. 

After \textbf{Step 1}, node $M$ waits for $t_{w}$ seconds to ensure that all these future transactions can arrive at corresponding nodes successfully before \textbf{Step 2} is executed. 

\textbf{Step 2.} Node $M$ sends $TX_M$ to all nodes $C$  excluding any  $A_k$ or node $B_k$ to transform these future transactions to pending transactions sequentially. In other words, $TX_M$ triggers the exchange of $TX_C^1$ along with $TX_M$ first. When receiving $TX_M$ and $TX_C^1$, $TX_C^2 $ becomes a pending transaction that will be forwarded to other nodes. 
Based on the discussion of the single link inference model, the link connecting $A_k$ and $B_k$ exists if $TX_B^k$ can be forwarded to node $M$.

\begin{figure}[tp]
\centering
\includegraphics[width=\linewidth,height=5.3cm]{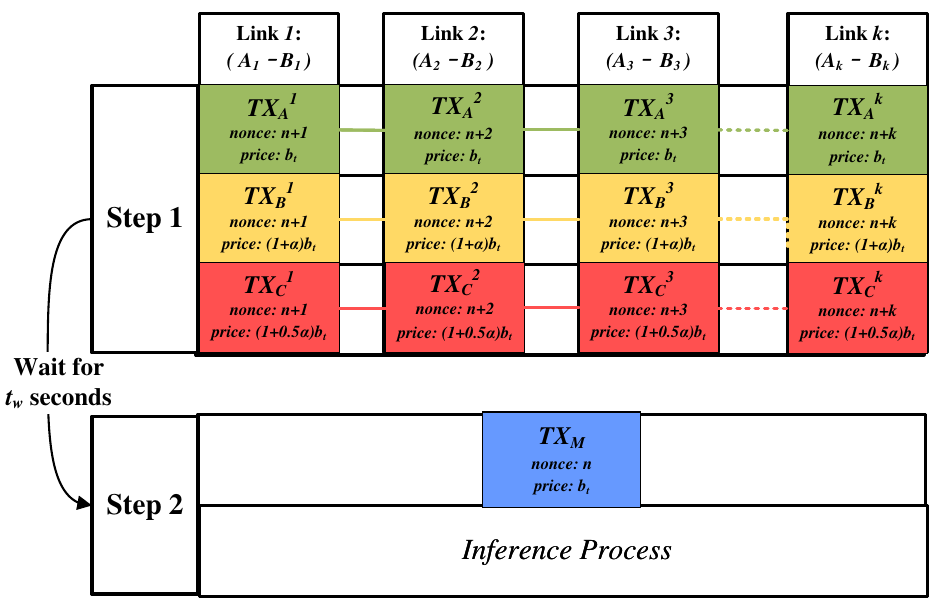}
\caption{Multi-link inference model: at Step 1, three marked transactions with the same nonce but different effective prices in each column are used to infer a link. $k$ marked transactions with continuous nonces and the same effective price in each row are sent to different nodes; at Step 2, $TX_M$ is sent to nodes $C$, which are the nodes in the network except $A_k$ and $B_k$.}
%(a) Parallel inference model to infer $k$ links with a single $TX_M$; (b) An example to infer 2 links after parallel step 4.}}
%%%(a) three marked transactions with same nonces but different effective prices in each column are used to infer a link, $k$ marked transactions with continuous nonces and same effective price in each row are sent to different nodes in parallel at each parallel step, and $TX_M$ is sent to node(s) $C$ at step 4, which are the nodes on the network beside for $k$ node $A$ and $k$ node $B$; (b) a link between node $A1$ and node $B1$ can be inferred if node $M$ receives $TX_{B1}$, and a link between node $A2$ and node $B2$ can be inferred if node $M$ receives $TX_{B2}$. }}
\label{fig3}
\end{figure}
To have a holistic overview of the multi-link inference model, we present its process in Fig.~\ref{fig3}. In \textbf{Step 1}, $K$ sets of marked transactions are distributed to corresponding nodes as future transactions. In \textbf{Step 2}, $TX_M$ is sent out to trigger the exchange of these marked transactions so that node $M$ can make link inference based on whether $TX_B^k$ can be forwarded back to $M$. 
Here, the effective price of these marked transactions is set the same as that in the single link inference model so that the forward of $TX_B^k$ is isolated except the link connecting $A_k$ and $B_k$.

%, consider that there is a single $TX_M$ with nonce $n$ and effective price $b_t$ and $k$ sets of marked transactions with the same account as $TX_M$ to infer $k$ links, specifically from link $(A1, B1)$ to $(Ak, Bk)$. The $i^{th}$ set of marked transactions to infer link $(Ai, Bi)$ are $TX_{Ai}$, $TX_{Bi}$, and $TX_{Ci}$ with the same nonce $n+i$ but different effective prices (i.e.,, respectively). Based on $k$ sets of marked transactions and a single $TX_M$, node $M$ works in five steps to infer $k$ links in parallel (5-step parallel inference task):

\noindent{\bf Improvement of Multi-link Inference.}
How much cost can be reduced by the multi-link inference model is dependent on the value of $K$. In theory, the maximum value of $K$ is $15$ because by default each node can store up to 16 pending transactions generated by the same account \cite{b28},  including  a single $TX_M$ with nonce $n$  and  $15$ marked transactions with nonce from $n+1$ to $n+15$. In this case, the multi-link inference model can roughly reduce 25\% of marked transactions and $\frac{14}{15}$ transaction fees for issuing $TX_M$ to measure a single link. 

%each node needs to store $k+1$ pending transactions and) with the same account after parallel step 4 but  Therefore, in the multi-link inference model, the number of marked transactions that need to pay the transaction fees is thereby from $l(l-1)$ to $\frac{{l(l - 1)}}{2} + \frac{{l(l - 1)}}{{2k}}=\frac{{8l(l - 1)}}{{15}}$ at most. 

%In the link inference model, a single $TX_M$ is solely utilized for inferring a link, and as a result, the number of marked transactions that need to pay the transaction fees is $l(l-1)$. In the parallel inference model, a single $TX_M$ is utilized for inferring $k$ links in parallel, and the number of marked transactions that need to pay the transaction fees is thereby decreased from $l(l-1)$ to $\frac{{l(l - 1)}}{2} + \frac{{l(l - 1)}}{{2k}}$. 

\subsection{Inference Failure Analysis}

%In the practical network, the multi-link inference model, which is an extension of the single link inference model, may encounter failed inference cases that affect the model's precision and recall. For convenience, we analyze these cases based on the multi-link inference model with $k=1$, i.e., the single link inference model.

There exist exceptional cases in which our link inference model fails to correctly infer the existence of a link. In this subsection, we describe three such  cases and also explain that the occurrence probability of these cases is very low implying that DEthna can achieve high inference accuracy. 

\subsubsection{Exceptional Case 1}%{\emph{Case 1}}: 
In the design of Geth, which is one of the original implementations of the Ethereum protocol, with version before v1.10.18,  a node will discard newer future transactions with a higher priority if its Queue is full, which can store at most 1,024 future transactions. Therefore, it is possible that target nodes $A$ and $B$ discard marked transactions before link inference is completed if their Queues are full. In this case,  DEthna fails to infer the existence of the link connecting $A$ and $B$. 

In reality, this exceptional case almost does not impair the effectiveness of DEthna for two reasons. First,  discarding newer transactions with a higher priority is a glitch in the design of Geth \cite{b17}. Geth has addressed the glitch of discarding newer high-priority transactions since v1.10.18 \cite{b34}, prioritizing staler transactions when the Queue is full.
%Since v1.10.18 \cite{b34}, Geth has revised its mechanism to discard staler transactions  first when Queue is full. 
Second, executing DEthna during off-peak hours reduces the likelihood of a Queue being fully occupied by valid transactions.

%Second, if DEthna is executed in off-peak hours, it is unlikely that a Queue will be fully occupied by valid transactions. 

%to ensure that as many nodes as possible adopt Geth after v1.10.18, our inference experiment was conducted half a year after our proposal was adopted.

%The first failed inference case is from an unreasonable design in the Ethereum software Geth that prevents the future transactions ($TX_A$, $TX_B$, and $TX_C$) from being stored at step 1, 2, and 3. For the node adopting Geth ,According to this design, $TX_A$, $TX_B$, and $TX_C$ are discarded by node $A$, node $B$, and node(s) $C$, and the last two steps in the link inference model will also fail. %{\bf YP: what is the meaning of younger age or older age? is it Nonce?}

%{\bf YP: better to use cite command}
%To address this issue, we engaged with the core developers of the Ethereum community and submitted a proposal \cite{b17}  that redesigns Geth to prioritize discarding the future transaction with older age, since it is more likely that the future transaction with younger age can wait for the pending transaction issued by the same account and then be moved to Pending. Our proposal was adopted and implemented in Geth after v1.10.18 \cite{b34}. Moreover, to ensure that as many nodes as possible adopt Geth after v1.10.18, our inference experiment was conducted half a year after our proposal was adopted.
%{\bf YP: remove the word Shenzhen or Shenzhen University. Is it possible to infer authors by tracking the version change of Geth?}

\subsubsection{Exceptional Case 2}
%\textbf{\emph{Case 2}}: 
The second exceptional case is rooted in node churn in the Ethereum network. In Ethereum, the churn of target nodes, \emph{i.e.}, nodes $A$ and $B$ to be measured, can fail link inference due to the lack of synchronization with the global state. In other words, it is possible that a newly arrival node or an existing node with its connection temporarily  lost  cannot complete its synchronization with the global state. Such nodes cannot verify and forward transactions due to the lack of necessary information about the state of the Ethereum network. It implies that the measurement of links connecting unsynchronized nodes fails. 

Fortunately, we design DEthna with the capability to identify and select synchronized nodes in Ethereum, which can effectively mitigate the chance of this exceptional case. In DEthna, node $M$ is  a synchronized node that can monitor and verify the transactions from other nodes in the network. Thus, only nodes that send verifiable transactions after connecting to node $M$ are identified by DEthna  as synchronized nodes. For unsynchronized nodes, DEthna waits until they finish synchronization before including them for topology discovery. 
\begin{figure}[tp]
\centering
\includegraphics[width=0.8\linewidth,height=3.5cm]{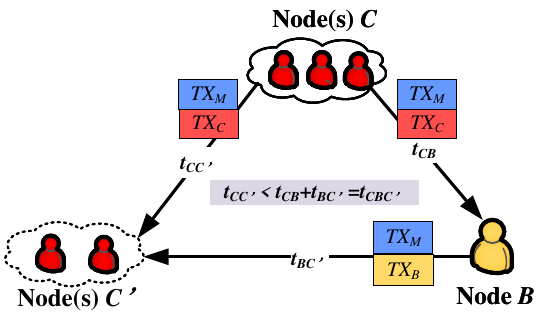}

\caption{The workflow when there exist nodes $C$’ not connecting with node $M$ can lead to the absence of $TX_C$. $TX_B$ is still isolated by nodes $C$’ when $TX_C$ arrives at node $C'$ earlier than $TX_B$.}
\label{fig2}
\end{figure}
\subsubsection{Exceptional Case 3}

%\textbf{\emph{Case 3}}: 

The last exceptional case is from the limited  connection capacity of the measurement node $M$. As presented in Fig.~\ref{fig2}, there may exist nodes $C'$ which are invisible to node $M$. 
Recall that our link inference model is based on the property that the forward of $TX_B$ from $B$ to other nodes except $A$ is isolated.   Due to the existence of nodes  $C'$, our link inference may fail because $TX_B$ can be relayed by $C'$ from node $B$ to node $A$. Once node $A$ receives $TX_B$ from $C'$, it can further forward it to $M$, making $M$ incorrectly infer  a direct link between $A$ and $B$. 

%The reason is that $TX_M$ is only sent to node(s) $C$ by $M$. Without $TX_M$. In other words,  when $TX_C$ is forwarded between nodes, $TX_B$ is still a future transaction that will not be forwarded to other nodes until $TX_M$ together with $TX_C$ can be forwarded to node $B$. 
In fact, DEthna can mitigate the disturbance of nodes $C'$ on link inference because it is more likely that $TX_C$ can be propagated to nodes $C'$ before $TX_B$, and thus isolate the forward of $TX_B$ from $B$ to $C'$. The reason is that $TX_M$ is only sent to nodes $C$ by $M$. In other words,  when $TX_C$ is forwarded between nodes, $TX_B$ is still a future transaction that will not be forwarded to other nodes until $TX_M$ together with $TX_C$ can be forwarded to node $B$.  Then, it takes at least a further forward action from $B$ to $C'$, which likely consumes a longer time. In contrast, if $TX_C$ can arrive at $C'$ before $TX_B$, it can isolate the forward of $TX_B$ from $B$  to  $C'$. 

To better understand how DEthna isolates $TX_B$ from $B$ to $C'$, let $t_{CC'}$ and $t_{CBC'}=t_{CB}+t_{BC'}$ denote the consumed propagation time for $C\rightarrow C'$ and  $C\rightarrow B\rightarrow C'$, respectively, which have been highlighted in Fig.~\ref{fig2}. Note that $TX_B$ will not be forwarded to $C'$, \emph{i.e.}, isolated by $TX_C$, as long as $t_{CC'}<t_{CBC'}$, which is guaranteed by \textbf{P1} of our link inference model and the triangle inequality between $t_{CC'}$, $t_{BC'}$ and $t_{CB}$.

The influence of $C'$ can be further alleviated by DEthna which can expand the connection capacity of node $M$ with multiple probing nodes. If $C$  can include most nodes in Ethereum, the influence of $C'$ can be ignored. In the next section, we will introduce a lightweight distributed implementation method of our multi-link inference model, through which we can significantly expand the connection capacity of node $M$ without incurring a heavy deployment cost.

\section{Distributed Implementation}\label{linkParallel}

%In this section, we first configure the parameters (including $n$, $b_t$, $\alpha$, and $t_{w}$) in the link inference model to finish the inference. 

In this section, we propose a low-cost approach to deploying multiple measurement nodes for Ethereum network discovery in a parallel and distributed mode to: 1) overcome the limited connection capacity of a single measurement node; and 2) expand the population of $C$ to avoid the occurrences of \emph{Exceptional Case 3}.

It is known that Ethereum nodes are scattered worldwide residing in different ISPs \cite{b33}. It is possible that  connections between the measurement node and other nodes are not established  due to traffic blocking from ISPs, which can hinder the efficiency of a single measurement node in discovering a large-scale Ethereum network. Intuitively speaking, this challenge can be overcome if multiple measurement nodes in different ISPs can be deployed to cooperatively measure the Ethereum network topology. However, a straightforward multi-node measurement architecture can incur significant and even unaffordable deployment costs. For example, the cost of a single high-performance node Alibaba cloud server (equipped with 8-core CPU, 16GB RAM, 1TB SSD, and 32 Mbit/s bandwidth) is \$180   for a week (specific cost information can be referred at https://www.alibabacloud.com/).%Specific cost information can be found on the official Alibaba Cloud website at.

\begin{figure}[tp]
\centering
\includegraphics[width=\linewidth,height=4.3cm]{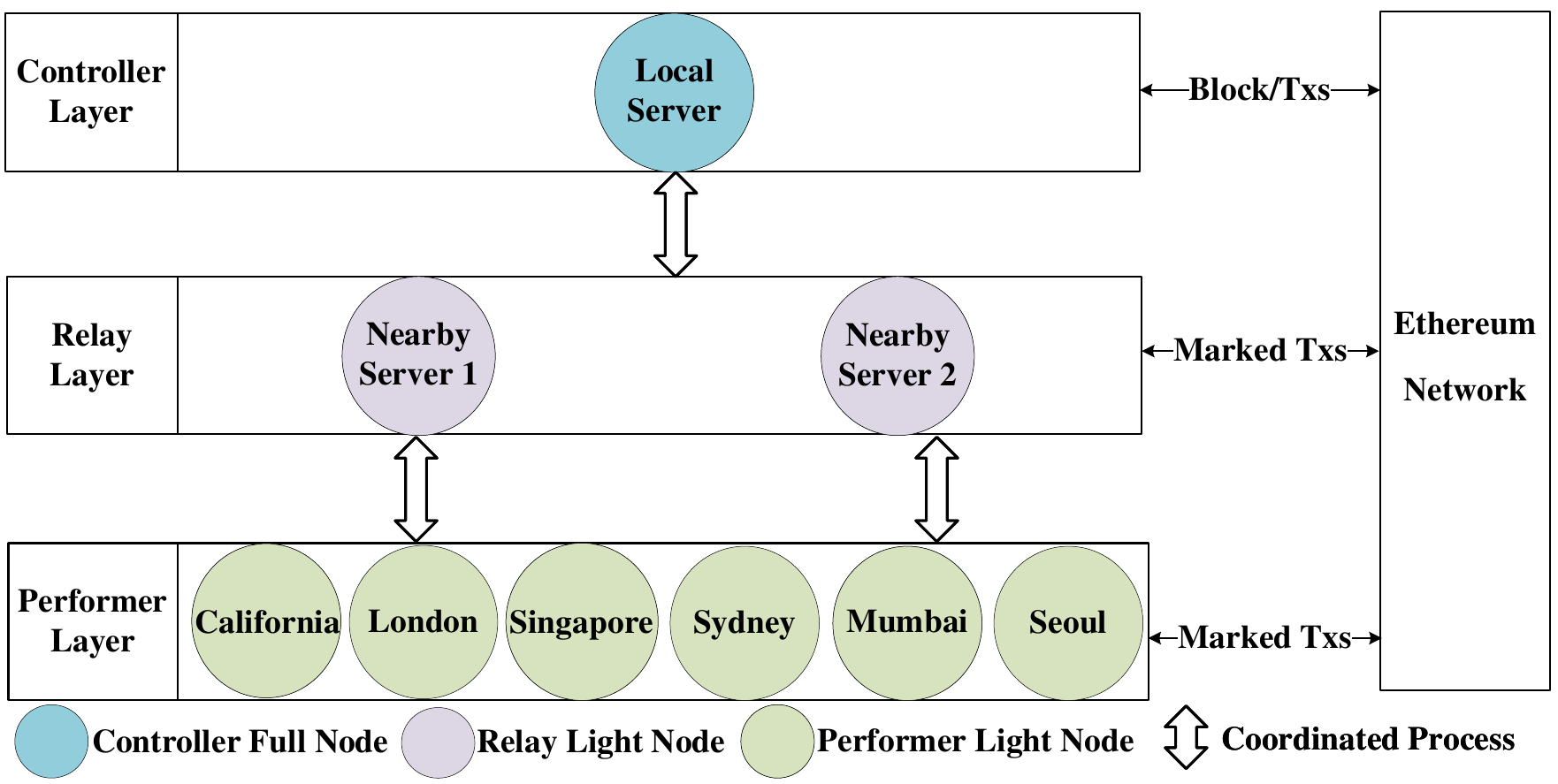}
\caption{The distributed measurement architecture consisting of 9 modified Ethereum nodes located around the world to act as node $M$.}
\label{fig3link}
\end{figure}

To minimize the deployment cost of multiple measurement nodes, we propose to only offload network discovery workload to rented cloud servers, but locally retain memory and computation resource consuming workload as much as possible.  To be exact, our design has three layers as shown in Fig.~\ref{fig3link}, \emph{i.e.}, the performer layer, the controller layer, and the relay layer,  and the functionality of each layer is described as follows:

%To solve the two problems of setting up multiple measurement nodes to play node $M$, we design a distributed measurement architecture, as shown in Fig.\ref{fig3link}. The architecture consists of three layers: Control Layer, Relay Layer, and Perform Layer. Within these layers, three types of modified measurement nodes are strategically deployed worldwide, including the control full node, relay light node, and perform light node. And their specific functions are elaborated as follows: 
\begin{itemize}
    \item \textbf{Performer Layer.} Multiple nodes are deployed in the performer layer by renting low-performance cloud servers geographically located in different regions over the world. Each measurement node pretends a standard Ethereum node to communicate with others in the network.  To reduce the consumption of cloud resources, measurement nodes will not spend resources on maintaining the global state of Ethereum, but mainly focus on two key processes. One is discovering and connecting Ethereum nodes in the network by conducting the routing table download algorithm in \cite{b11}; the other is executing link inference tasks from the controller layer to send and monitor marked transactions according to the link inference model.

   % The perform light nodes are designed to operate efficiently on low-performance servers by not maintaining the global state but  This design helps to reduce costs associated with renting cloud servers. Also,  thereby facilitating the discovery of a complete network topology. 

\item \textbf{Relay Layer.} Nodes in the relay layer are also rented from cloud providers that are strategically located geographically near the node in the controller layer with two roles. On the one hand, relay nodes perform the network discovery function and link inference tasks like nodes in the performer layer. On the other hand, relay nodes act as intermediaries for message forwarding between the controller layer and the performer layer if any performer layer node fails to directly communicate with the controller layer node due to traffic blocking by ISPs. 

%The relay light nodes serve a dual purpose. One is they have the same function as the perform light nodes. Another is they This is necessary because the control full node is deployed on our local server in Shenzhen and direct connections between the control full node and perform light nodes may not be feasible due to .  

\item \textbf{Controller Layer.} The node in the controller layer is implemented on a local high-performance server not rented from cloud providers, which is used  
%The control full node  serves 
as the central control point of the distributed network measurement task. It is responsible for generating link inference tasks and maintaining the global state of Ethereum by exchanging blocks and transactions with the Ethereum network. Additionally, it coordinates the parallel link inference measurement tasks executed on multiple measurement nodes by sharing the information of connecting nodes to eliminate duplicated link measurement tasks. 
 %in the bottom layers through coordinated processes.  The coordinated processes ensure efficient cooperation and synchronization among the modified measurement nodes, enabling them to perform the same inference task effectively and in a coordinated manner.
\end{itemize}

{Based on the three-layer design,  the cost of a measurement node in the relay or performer layer is thereby decreased from \$180 to \$30 by renting a low-performance cloud server (2-core CPU, 4GB RAM, 40GB SSD, 8Mbit/s) for a week.}

\section{Experiment Study} \label{experiment}
In this section, we evaluate the performance of DEthna on Goerli, the most popular Ethereum test network. Then, we analyze the network characteristics of Ethereum based on the network topology discovered by DEthna.

\subsection{Settings of DEthna Evaluation}\label{experiemtn1}

%we first introduce the detailed parameter configuration in the link inference model. \subsection{Detailed Parameter Configuration}\label{linkb}
%\subsubsection{System Settings}
\noindent{\bf Key Parameter Settings.}
In our experiments, we control node $M$ to discover the network topology by setting four key parameters as follows: 1) $n$ ($TX_M$’s $nonce$) is set according to node $M$'s local database. Here, $M$ is a full Ethereum node that can synchronize the global state. 
2) $b_t$ ($TX_M$’s effective price) is set to match that of the $1429^{th}$ transaction (ranked by a descending order of effective price of transactions) in the transaction pool of node $M$  by cloning $gasFeeCap$ and $gasTipCap$ of the ${1429^{th}}$ transaction. 
3) $\alpha$ (Replacement Rate) is set to $0.1$, which is the default replacement rate widely used by Ethereum software like Geth, Erigon, Besu, and Nethermind.  
4) $t_{w}$ (Waiting Interval) is the time required for node $M$ to upload all future transactions ($TX_A, TX_B$, and $TX_C$) at Step 1. In our experiments, this interval is set to 3.5 seconds, allowing node $M$ to use multiple accounts to infer 160 links in parallel.

Note that  DEthna sets $b_t$  as the minimum value to complete link inference.  On the one hand, the transaction pool of a node is limited. 
Only the transaction with its effective price higher than a threshold can be stored and forwarded over the network when competing with other transactions. On the other hand, if $b_t$ is too high,  marked transactions issued by $M$ will be packed in the block instantly without being forwarded over the network until the link inference process is completed. {In Ethereum, there is a fixed gas limit of 30,000,000 to restrict the number of transactions packed in a block. %, \emph{i.e.}, the total gas consumed by transactions in a block cannot be larger than 30,000,000. 
The minimal gas consumed by a transaction is 21,000. A block can thereby contain $30,000,000/21,000 \approx 1,429$ transactions at most. Thereby, the price of the $1,429$-th transaction is the minimum value to guarantee the propagation of our transactions.}
%discussed in the last section, pending nonce is the number of transactions that are issued by account $X$ and have been committed in the global state. Because node $M$ is a full Ethereum node that fully synchronizes the global state, node $M$ can access its local database to ensure $n$.

%: $b_t$ should be configured properly. On the one hand, $b_t$ should be configured high enough, since the space of the transaction pool is limited and only the transaction with a higher effective price can be stored and forwarded over the network when the transaction pool is full.  Therefore, a proper $b_t$ that ensures the marked transactions cannot be packed in the next block is configured as follows.

%\footnote{The current block interval in Ethereum 2.0 is fixed by 12 seconds. If the marked transactions are not packed in the next block, there are more than 12 seconds to perform the inference. And the duration is long enough to finish the inference task. }

%For the transactions in node $M$’s transaction pool, node $M$ sorts them in a descending order by their effective price, and then configure $b_t$ ). %{\bf YP: it is uncertain how many gas a transaction will consume. why do not we copy the gas fee of the transaction which just consume the 30,000,000-th gas}

%\footnote{Sometimes the number of transactions in the transaction pool is less than 1430. In this case, to guarantee as much inference time as possible, node $M$ selects the last transaction as the reference and generates the four marked transactions immediately to start the inference after receiving a new block.}
\begin{table}[] \centering
	\caption{Link inference performance of different schemes}
\begin{tabular}{|l|c|c|c|c|c|}
\hline
\diagbox{\textbf{Scheme}}{\textbf{Metrics}}  & \textbf{Precision} & \textbf{Recall} & \textbf{F1-Score} \\ \hline
K-Bucket          & 0.026     & 0.061  & 0.036    \\ \hline
Basic TopoShot    & 0.977     & 0.325  & 0.488    \\ \hline
Improved TopoShot & 0.980     & 0.536  & 0.693    \\ \hline
DEthna            & 0.983     & 0.889  & 0.934    \\ \hline

\end{tabular}

\label{pre}
\end{table}

\noindent{\bf Ground Truth Collection.}
To collect the ground truth knowledge for evaluating  inference performance of DEthna, we run a standard full Ethereum node with the typical configuration in \cite{b28} on Goerli. This Ethereum node plays the role of node $A$  in DEthna by  randomly connecting to other nodes on Goerli and synchronizing the global state by exchanging blocks and transactions from Goerli. It  records all connected nodes during the period from November 17, 2022 to December 1, 2022. 

We run DEthna on November 19, 2022 
to infer links  connecting  with the standard full Ethereum node every 8 hours. %and counted the precision and recall of DEthna by
Based on collected ground truth knowledge,  we can thereby evaluate the link inference performance.

\noindent{\bf Baselines.}
Other than DEthna,  we implement three  link inference schemes  on Goerli as baselines. % to infer the links of the standard full Ethereum node: 
The first one is the \textbf{k-bucket} scheme that uses the node information stored in the k-bucket data structure \cite{b11, b12, b13} to infer links. The second one is a basic \textbf{TopoShot} scheme \cite{b14} that  infers links by generating  marked transactions with a single measurement node. 
However, a notable drawback of this scheme  is  that a large number of invalid transactions will be generated to evict valid transactions in a  node's transaction pool.   The last one is the \textbf{improved TopoShot} scheme, which implements the original TopoShot scheme based on our distributed architecture so that the link inference process can be accelerated with  multiple measurement nodes.
%that uses our DEthna architecture that consists of multiple measurement nodes to perform the similar behavior in the basic TopoShot, 
In our experiments, we implement these three baselines every 8 hours on 
  November 22, 28 and December 1, 2022, respectively. 

%We can see that our DEthna stands out with its superior precision (0.97) and recall (0.89) metrics among the evaluated schemes. 

\subsection{Experimental Results of DEthna Evaluation}
\noindent{\bf Comparing Link Inference Performance.}
The experimental results are presented in Table \ref{pre} by comparing precision, recall and F1-Score achieved by different link inference models. From Table \ref{pre}, we can see that: 1) DEthna is the best one achieving the highest precision, recall and F1-Score; 2) In particular, the recall of DEthna is much higher than baselines implying that a large number of links in the Ethereum network cannot be discovered by  baselines; 3) The k-bucket scheme is the worst one  with very low precision and recall. The reason lies in that most of the remote nodes in the k-bucket are not really connected for establishing the links;
%This scheme performs the worst in terms of precision and recall among the four schemes evaluated, . For the remaining three schemes, their precisions are high and close to each other, around 0.97. However, a single measurement node used in the basic TopoShot would miss discovering a large of Ethereum nodes on the network, and its malicious behavior that affects the recall has been limited by a countermeasure proposed by the Ethereum community \cite{b15}, the recall of the basic TopoShot is thereby only around 0.325. 
4) Improved TopoShot is the second best one, and this result also indicates that the multi-node discovery architecture proposed by us can effectively improve link inference accuracy.  
%improved to around 0.536 by introducing our DEthna architecture, it is still limited by the countermeasure. And the recall of our DEthna is larger than that of other schemes significantly, around 0.890. That is, our DEthna would miss detecting 11\% of links over Goerli, since some Ethereum nodes only joined Goerli but had not fully synchronized the global state to forward the transactions or used a replacement rate $\alpha$ that is larger than 0.1. Therefore, with the help of a precision of 0.97 and a recall of 0.89, our DEthna can discover a relatively complete network topology of Ethereum. 

%\textbf{\emph{Network Size}}: 

\noindent{\bf Comparing Scalability of Link Inference Schemes.}
To validate the supreme scalability of DEthna using multiple nodes for network discovery, we compare the network size discovered by each link inference scheme versus measurement duration.  
The network size is defined as the number of Ethereum nodes connecting to measurement nodes in DEthna. 
The discovered network size is crucial for discovering the  network topology accurately and completely. We execute DEthna on Goerli to measure its network size for 48 hours (from November 3, 2022 to November 5, 2022). As a comparison, we also set up a single measurement node (used in basic TopoShot and executed on a high-performance Alibaba Cloud server) for discovering network topology over the same period. %It should be noted that unlike a standard Ethereum node, which is typically limited to connecting with a maximum of 50 other Ethereum nodes on the network \cite{b16}, both the measurement nodes used in DEthna and the single measurement node used in  the comparative experiment do not have such limit. 
%Unlike a standard Ethereum node, which is typically limited to connecting with up to 50 Ethereum nodes [16], the modified light nodes used in DEthna and the full Ethereum node used in the comparison experiment have no such limits. 
%That is, they are free to discover and connect to Ethereum nodes as much as possible. 
%\footnote{The Alibaba Cloud server was located in California and had a high hardware configuration with 1TB SSD, 20 Mbit/s bandwidth, 16GB RAM, and 8-core CPU.}

\begin{figure}[tp]
\centering
\includegraphics[width=\linewidth, height=3.3cm]{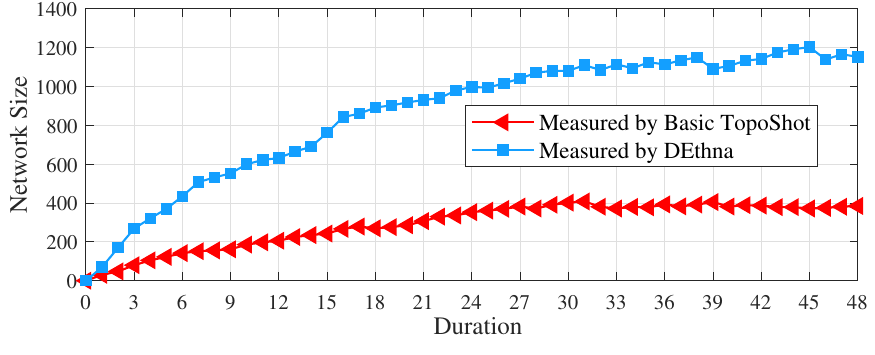}
\caption{The network size of Goerli measured by DEthna and Basic TopoShot.}
\label{networksize}
\end{figure}

The discovered network sizes of Goerli measured by the multi-node architecture of DEthna and the single-node architecture used in basic TopoShot are shown in Fig.~\ref{networksize}. We can see that the network size increases with respect to the elapsed measurement time, and becomes relatively stable after 30 hours. %, since both DEthna and the single measurement node need a certain amount of time to discover the Ethereum nodes on Goerli. 
DEthna can  discover about 1,150 nodes in Goerli for 30 hours.  
In comparison, the inference model with a single measurement node used in basic TopoShot can only discover about 400 Ethereum nodes over the same period. 
Therefore, the distributed measurement architecture of DEthna can greatly improve its network discovery efficiency and accuracy.

%to connect to other Ethereum nodes in the network before it starts to discover the network topology. Meanwhile, DEthna discovered and connected to around 1, but the single measurement node only discovered and connected to around 400 Ethereum nodes. If we only set up a single measurement node to play node $M$, around 2/3 of links cannot be inferred. 

%\begin{figure}[tp]
%\centering
%\includegraphics[width=9cm,height=3.8cm]{Fig6_a.pdf}
%\caption{The precision and recall of different schemes to %infer links on Goerli.}
%\label{papers}
%\end{figure}

\begin{table}[]
	\caption{Basic characteristics of the Goerli testnet}

\begin{tabular}{|c|p{0.9cm}<{\centering}|p{0.6cm}<{\centering}|p{0.8cm}<{\centering}|p{0.9cm}<{\centering}|p{0.95cm}<{\centering}|}
\hline
\diagbox[width=2.4cm]{\textbf{Network}}{\textbf{Statistics}}     & \begin{tabular}[c]{@{}c@{}}\textbf{Network}\\ \textbf{Size}\end{tabular} & \begin{tabular}[c]{@{}c@{}}\textbf{\# of} \\ \textbf{Links}\end{tabular} & \begin{tabular}[c]{@{}c@{}}\textbf{Average}\\ \textbf{Degree}\end{tabular} & \begin{tabular}[c]{@{}c@{}}\textbf{Average}\\ \textbf{Shortest}\\ \textbf{Path}\end{tabular} & \textbf{Diameter} \\ \hline
Goerli-0102 & 1,193                                                  & 10,552                                                 & 17.6                                                     & 2.69                                                              & 6        \\ \hline
Goerli-0116 & 1,265                                                  & 11,828                                                 & 18.7                                                     & 2.72                                                              & 6        \\ \hline
\end{tabular}

	\label{matrixcharacteristics}
\end{table}

\subsection{Network Characteristics Analysis}
Given the highly accurate link inference results of DEthna, we further analyze the characteristics of the Ethereum network based on which we can exploratively discuss how to improve the security and robustness of Ethereum. 

%{\bf YP: Table II is out of boundary. It is necessary to adjust it to fit the column size. }
We execute DEthna on Goerli on January 2, 2023 and January 16, 2023, respectively,  to discover two slightly different network topologies, which are named Goerli-0102 and Goerli-0116. The network characteristics to be analyzed  include node degree distribution, broadcasting hops and network robustness. % to give a clear direction for the optimization of the Ethereum network. 

\noindent\textbf{{Basic Network Characteristics.}} Table~\ref{matrixcharacteristics} presents the summary of basic characteristics of the Goerli-0102 and Goerli-0116 networks in terms of network size (\emph{i.e.}, the number of nodes), link population, average degree, average of the shortest path length (between any two nodes) and the network diameter (defined as the maximum value of the shortest path length between any two nodes in the network). {From  Table~\ref{matrixcharacteristics}, we can find that the basic characteristics of  Goerli-0102 and Goerli-0116  are very similar with the network diameter equals to $6$, \emph{i.e.}, any node takes no more than 6 hops for broadcasting messages to all other nodes. This result verifies the conjecture on the Ethereum network in \cite{b16}.} Moreover, compared to \cite{b16}, our work takes one more step to further analyze the robustness and security of the Ethereum network hereafter.

%{\bf YP: removing the repetition of these numbers you have presented in Table II, we can save some space.} The result had excluded the modified nodes in DEthna and its links. For network size and edges, Goerli-0102 contains 1193 nodes and 10825 links among them, while Goerli-0116 contains 1265 nodes and 11828 links; for degree, the nodes on Goerli-0102 maintain an average of 17.6 neighbor nodes to exchange the blockchain messages, while the nodes on Goerli-0116 maintain an average of 18.7 neighbor nodes; for short path, defined as the distance between a pair of nodes, the nodes on Goerli-0102 require an average of 2.69 hops to broadcast a message to other nodes, while the nodes on Goerli-0116 require an average of 2.72 hops; for diameter, defined as the maximal distance between any pair of nodes, both nodes on Goerli-0102 and Goerli-0116 require at most 6 hops to broadcast a message to other nodes. By comparing the basic properties of the Goerli in different periods, we can find that the Goerli keep relatively stable within two weeks.

\begin{figure}[tp]
\centering
\includegraphics[width=\linewidth, height=3.0cm]{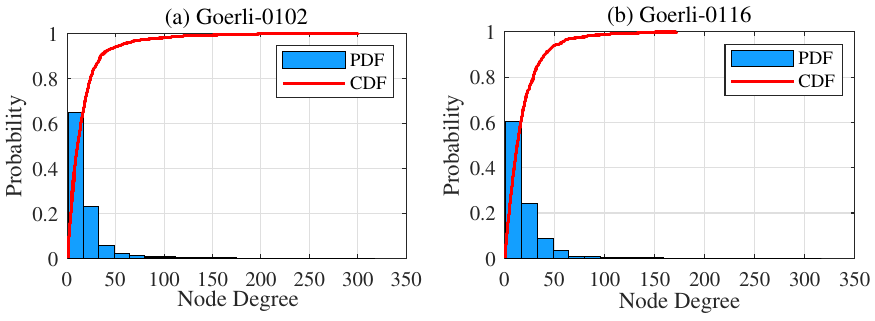}
\caption{The empirical PDF and CDF of node degree in two Goerli networks.}
\label{fig6con}
\end{figure}

\noindent\textbf{{Degree Distribution and Node Classification.}}
Compared to the average degree, degree distribution can provide a deeper understanding of node importance and network structures. 
%Node degree provides an intuitive measure of the importance of a node in a network, and the degree distribution is important for understanding the structural properties of the network.
Fig.~\ref{fig6con} presents the empirical probability density functions (PDF) and cumulative distribution functions (CDF) of node degrees in  Goerli-0102 and Goerli-0116, respectively. %Meanwhile, Fig.\ref{fig6con}(c) and Fig.\ref{fig6con}(d) show empirical  for the node degrees in the same two networks, Goerli-0102 and Goerli-0116.
From Fig.~\ref{fig6con}, we can see that the majority of nodes in the network have a degree  less than 50. Specifically, the degree of more than 93\% nodes %in Goerli-0102 and 93.4\% of nodes in Goerli-0116 have a degree of
is less than 50 in both networks. 

Based on  Fig.~\ref{fig6con}, we can classify all nodes into three types. 
%Furthermore, 62.3\% of nodes in Goerli-0102 and 62.6\% of nodes in Goerli-0116 have a degree of less than 17. That is, 93.8\% of nodes in Goerli-0102 and 93.4\% of nodes in Goerli-0116 are configured with 
In Ethereum network,  the default maximum degree of a node is 50 \cite{b16}. Each node at most establishes $50/3\approx 16$ outbound connections and $50*2/3\approx 34$  inbound connections  to prevent ``false friends" attacks. {It is also possible that the degree of a node exceeds 50 if the node targets to make more profit or conduct specific monitoring activities in Ethereum by locally modifying the default $50$ degree constraint. For example, \cite{b3} set up a monitoring node with a degree of around 1,000 in Ethereum to analyze the private behavior of miners.} 

%; 62.3\% of nodes in Goerli-0102 and 62.6\% of nodes in Goerli-0116, which have a degree less than 17, are behind NAT without a fixed port and publish IP so that they can only establish $50/3 \approx 16$ outbound connections, since each node establishes $x/3$ outbound connections and $2x/3$ inbound connections at most to prevent the “false friends” attack \cite{b36}. Meanwhile, 6.2\% of nodes in Goerli-0102 and 6.6\% of nodes in Goerli-0116 have a degree exceeding the default maximum degree of 50, by acting as super nodes, and the super nodes would accelerate the message propagation but may weaken the decentralization of the network. 

Based on node degrees, nodes can be classified into three types: low-degree nodes with a degree no more than $16$, ordinary nodes with a degree in $(16, 50]$, and super nodes with a degree greater than $50$. 
For low-degree nodes, it is likely that they are Ethereum nodes behind NAT or firewalls who can only establish outbound connections with other nodes. As a result, their degrees are no more than $16$. 

We will conduct more analysis for the three types of nodes separately to understand how each type of nodes affects the Ethereum network robustness and security.

%the analysis of the node degree distribution in the Goerli-0102 and Goerli-0116 networks suggests that these networks have a relatively sparse and decentralized structure with many nodes having only a few connections to other nodes. However, the presence of super nodes with degrees exceeding the default maximum degree of 50 highlights the potential for centralization and the need for careful monitoring to ensure the network remains decentralized and secure.

\begin{figure}[tp]
\centering
\includegraphics[width=\linewidth,height=3.0cm]{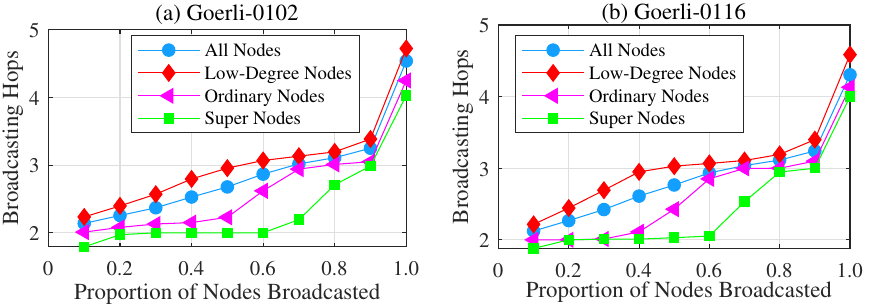}
\caption{Broadcasting hops for different nodes to broadcast the messages in Goerli-0102 and Goerli-0116.}
\label{fig7conaa}
\end{figure}
\noindent\textbf{{Broadcasting Hops.}} The number of broadcasting hops is the number of hops required for a node to broadcast messages (\emph{e.g.}, transactions and blocks) to reach all other nodes in the Ethereum network. 
It is a crucial factor  reflecting network efficiency and security \cite{b37, b38,dino}. To analyze different message broadcasting capabilities of different types of nodes, we show 
 the number of broadcasting hops for each type of nodes in Goerli-0102 and Goerli-0116, respectively, in Fig.~\ref{fig7conaa}.

From Fig.~\ref{fig7conaa}, for both Goerli-0102 and Goerli-0116 networks, we can observe that: %1) On average, it takes  no more than 5 hops for broadcasting messages to all other nodes.} 
%\textcolor{red}{ Chonghe: The boxplot has been removed from fig.7 so that fig.7 cannot reflect this conclusion. But it can be reflected from Table 2 (network diameter is 6). Does this sentence need to be deleted? or add the words "from Table 2"? 2) 
1) It takes fewer hops for super nodes to broadcast messages than ordinary and low-degree nodes.  2) The network topology of Goerli leads to ``long tail” broadcasting latencies because it takes about 3 hops to broadcast messages to about  90\% nodes, but takes more than 1 hop to further reach the remaining nodes in the network. % after they have broadcast messages to 90\% of the remaining nodes with 3.24 hops.
It is worth noting that broadcasting by super nodes cannot get rid of the ``long tail”  latency problem.

\noindent\textbf{{Network Robustness.}} 
Network robustness is the ability  to maintain network structural integrity and  functionality after being attacked or experiencing node failures. 
{It is particularly important for the Ethereum network to maintain a high level of security and stability to defend against various attacks, such as double spending attacks, 51\% attacks and DDoS attacks \cite{b48}.}
%operate effectively \cite{}. 
According to \cite{b39}, the network robustness can be evaluated by simulating attacks to remove  nodes from the network. Then the network robustness is measured by the size of the remaining largest connected component relative to its original size.

Intuitively speaking, the network robustness is negatively influenced by low-degree nodes.  To show this point, we generate Goerli-0102R and Goerli-0116R networks which are generated by removing low-degree nodes  from Goerli-0102 and Goerli-0116, respectively. 
Meanwhile, we generate two classic networks, called ER and BA, to quantify the influence of attacks on Goerli based networks. Both ER and BA are generated by NetworkX \cite{b40} with the same network size as Goerli based networks.  Links in ER and BA are generated by the Erdos-Renyi model \cite{b41} and the Barabasi-Alber model \cite{b42}, respectively.
Finally, we generate ER-0102, BA-0102, ER-0116, BA-0116, ER-0102R, BA-0102R,  ER-0116R and BA-0116R networks, to compare with Goerli based networks. 

\begin{figure}[tp]
\centering
\includegraphics[width=\linewidth,height=2.7cm]{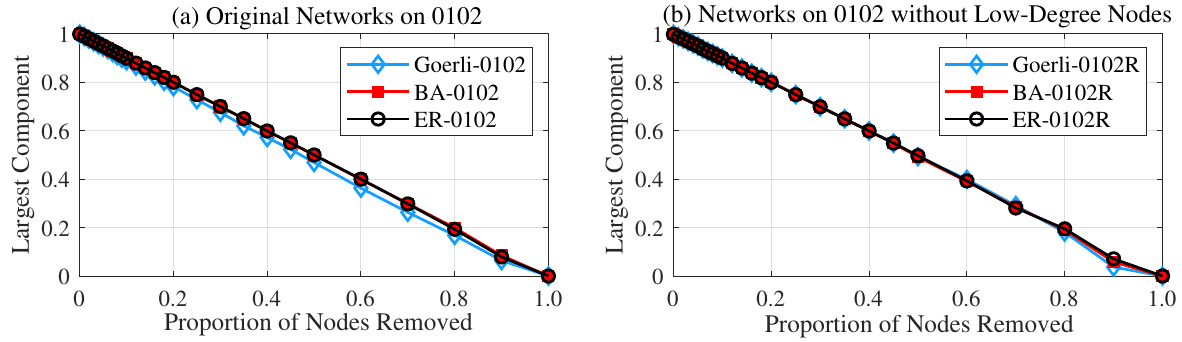}
\caption{Robustness of different networks against random attacks.}
\label{randomaaa}
\end{figure}

\begin{figure}[tp]
\centering
\includegraphics[width=\linewidth,height=2.7cm]{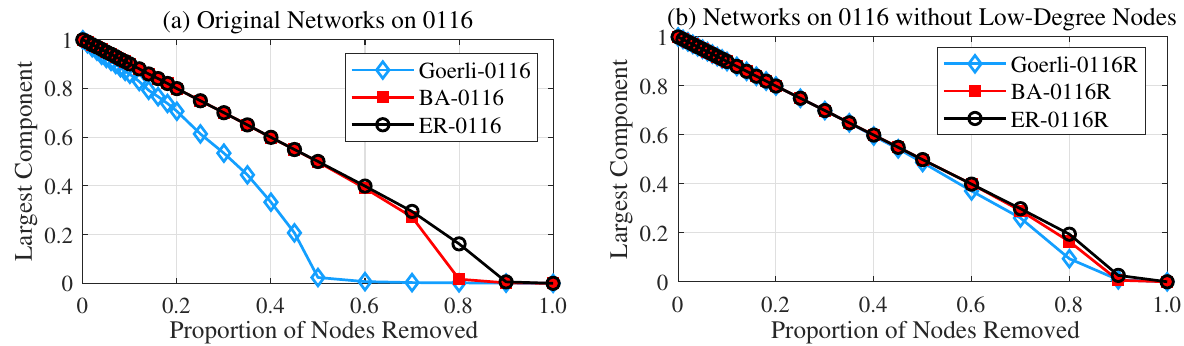}
\caption{Robustness of different networks against targeted attacks.}
\label{puraaa}
\end{figure}

%We implement two attacks to remove nodes which are the random attack and the targeted attack.
{We implement two types of attacks on different networks. From previous experimental results, we can observe that the two networks sampled on 0102 and 0116 are very similar to each other. Due to limited space, we just implement  random attacks on networks sampled on 0102 %(\emph{i.e.}, Goerli-0102, ER-0102, BA-0102, Goerli-0102R, ER-0102R, and BA-0102R) 
and targeted attacks on networks sampled on 0116, %based networks on ($i.e.$, Goerli-0116, ER-0116, BA-0115, Goerli-0116R, ER-0116R, and BA-0116R)
 respectively.} The random attack randomly selects a node from a network, while the targeted attack ranks nodes by a descending order of their degrees and removes the node with a larger degree with a higher priority. The experimental results are presented in Fig.~\ref{randomaaa} for the random attack and Fig.~\ref{puraaa} for the targeted attack, respectively. The x-axis represents the proportion of removed nodes while the y-axis represents the size of the largest connected component relative to the original network size after removing nodes, and  a larger area under the line indicates a more robust network.

%And then to quantify the robustness of Ethereum network under the random attacks, we conducted four sets of experiments by randomly removing the nodes from different networks, and the experimental results are shown in Fig.\ref{randomaaa}. 

%. Similarly, to quantify the robustness of Ethereum network under the targeted attacks, we conducted four sets of experiments by removing the nodes in order of decreasing node degree from different networks, and the experimental results are shown in Fig. \ref{puraaa}. 

%Each set of experiments contained three networks: Goerli, ER, and BA. Both ER and BA were generated by Network \cite{b40} according to the corresponding properties of Goerli: ER was the network with the same network size and links as Goerli following Erdos-Renyi model \cite{b41}, and BA was the network with the same network size and average node degree as Goerli following Barabasi-Alber model \cite{b42}. The specific networks in each set of experiments were given as follows: Goerli-0102, ER-0102, and BA-0102 in the first set; Goerli-0116, ER-0116, and BA-0116 in the second set; Goerli-0102’, ER-0102’, and BA-0102’ in the third set; and Goerli-0116’, ER-0116’, and BA-0116’ in the last set. Similarly, to quantify the robustness of Ethereum network under the targeted attacks, we conducted four sets of experiments by removing the nodes in order of decreasing node degree from different networks, and the experimental results are shown in Fig.\ref{puraaa}. 

From Fig.~\ref{randomaaa} and Fig.~\ref{puraaa}, we can conclude that: 1) Goerli based networks are robust to random attack because the relative network size is very close to that of BA or ER based networks after removing nodes. 2) Goerli based networks are vulnerable to the targeted attack, evidenced by  the gap between Goerli based network size and BA/ER based network size after removing a certain fraction of large-degree nodes. 3) The robustness of the Goerli network can be significantly improved if we can increase the degree of low-degree nodes because the curves of Goerli-0116R (which are generated by removing low-degree nodes from Goerli) are very close to these of BA/ER based networks. 

In summary, our robustness study reveals that the robustness of Ethereum is impaired by low-degree nodes, which are likely nodes behind NAT or firewalls. {There are two effective approaches to enhancing Ethereum network robustness. The first one is to expand the degree of these low-degree nodes. The second one is to make the link inference schemes unavailable to make it difficult for attackers to detect low-degree nodes. %For example, each node can configure its replacement rate $\alpha$ randomly.
}

\section{Conclusion}\label{conclusion}
Discovering Ethereum network topology is vital for improving network efficiency, scalability, and security. 
In this work, we propose  DEthna, a novel scheme to discover Ethereum network topology by generating marked transactions. Compared with existing schemes, DEthna is of the following advantages: 1) DEthna is friendly for implementation which averts flooding  the Ethereum network with invalid transactions; 2) The link inference accuracy of DEthna is much higher than existing schemes; 3) DEthna is efficient which can leverage multiple decentralized probing nodes to discover network topology in a parallel mode. By measuring Goerli with DEthna, we further analyze Ethereum network characteristics and reveal that Ethereum network robustness is compromised by low-degree nodes. 
Our initial research provides a better understanding of the Ethereum network, and lays the foundation to optimize it. For example, based on the  network topology measured by DEthna, in the  future we can study how to design a more advanced message propagation protocol to speed up the message propagation in the Ethereum network.
%{\bf YP: what is the future work? Can we outlook and briefly discuss what is the future work reseachers can do following our work? }

%A link inference model is proposed to infer a link between two nodes in Ethereum network by setting up a measurement node to perform a 5-step inference task. Based on the link inference model, a distributed measurement architecture consisting of multiple measurement nodes is further designed to connect more Ethereum nodes to discover a complete network topology. Compared to the existing schemes, DEthna achieves both high precision and recall. DEthna discover the topology of Goerli in two periods, which show more than half of NAT nodes in the network would lead to the “long tail” latencies to cover the entire network and a weak robustness to defend the targeted attack. 
%work %\section*{Acknowledgment}

%The preferred spelling of the word ``acknowledgment'' in America is without 
%an ``e'' after the ``g''. Avoid the stilted expression ``one of us (R. B. 
%G.) thanks $\ldots$''. Instead, try ``R. B. G. thanks$\ldots$''. Put sponsor 
%acknowledgments in the unnumbered footnote on the first page.

\clearpage

\vspace{12pt}

\bibliographystyle{IEEEtran}
%\bibliography{references}{}
%\bibliographystyle{splncs04} %这是一个样式文件
\bibliography{DEthna} %其中reference为reference.bib文件。
\end{document}